\documentclass[aps,prc,preprint,groupedaddress,showpacs,floatfix]{revtex4-1}
\usepackage{amssymb}
\usepackage{amsmath,bm}
\usepackage[normalem]{ulem}
\usepackage[dvips]{color}
\usepackage{booktabs}
\usepackage{graphicx}
\usepackage{float}
\usepackage{multirow}
\usepackage{lineno}
\usepackage[normalem]{ulem}
\usepackage[usenames, dvipsnames]{xcolor}
\usepackage{tensor}
\usepackage[utf8]{inputenc}
\usepackage{subfigure}
\newcommand{\ssr}{$\rm \frac{N_{^3_{\Lambda}H}/N_\Lambda}{N_{^3He}/N_p}$~}
\newcommand{\sr}{$\rm \frac{N_{^3_{\Lambda}H}}{N_\Lambda N_ d}$~}
\newcommand{\hyt}{$\rm ^3_{\Lambda}H$~}
\newcommand{\he}{$\rm ^3He$~}

\begin{document}

\title{Yield ratio of hypertriton to light nuclei in heavy-ion collisions from $\rm \sqrt{s_{NN}}$ = 4.9 GeV to 2.76 TeV}
\author{Tianhao Shao$^{1,2,3}$}\thanks{shaotianhao@sinap.ac.cn}
\author{Jinhui Chen$^{1}$}\thanks{chenjinhui@fudan.edu.cn}
\author{Che Ming Ko$^{4}$}\thanks{ko@comp.tamu.edu}
\author{Kai-Jia Sun$^{4}$}\thanks{sunkaijiaxn@gmail.com}
\author{Zhangbu Xu$^{5}$}\thanks{xzb@bnl.gov}

\affiliation{$^1$ Key Laboratory of Nuclear Physics and Ion-beam Application (MOE), Institute of Modern Physics, Fudan University, Shanghai 200433, China}
\affiliation{$^2$ Shanghai Institute of Applied Physics, Chinese Academy of Science, Shanghai 201800, China}
\affiliation{$^3$ University of Chinese Academy of Science, Beijing 100049, China}
\affiliation{$^4$ Cyclotron Institute and Department of Physics and Astronomy, Texas A\&M University, College Station, Texas 77843, USA}
\affiliation{$^5$ Brookhaven National Laboratory, Upton, New York 11973, USA}
\date{\today}

\begin{abstract}
We resolve the difference in the yield ratio $\rm S_3$ = \ssr measured in Au+Au collisions at $\rm \sqrt{s_{NN}}$ = 200 GeV and in Pb-Pb collisions at $\rm \sqrt{s_{NN}}$ = 2.76 TeV by adopting a different treatment of the weak decay contribution to the proton yield in Au+Au collisions at $\rm \sqrt{s_{NN}}$ = 200 GeV. We then use the coalescence model to extract information on the $\Lambda$ and nucleon density fluctuations at the kinetic freeze-out of heavy ion collisions. We also show from available experimental data that the yield ratio $\rm S_2$ = \sr is a more promising observable than $\rm S_3$ for probing the local baryon-strangeness correlation in the produced medium.
\end{abstract}
\pacs{25.75.-q, 25.75.Dw}
\maketitle

\section{Introduction}

The correlation coefficient $\rm C_{BS}=-3\frac{\langle BS\rangle-\langle B\rangle\langle S\rangle}{\langle S^2\rangle-\langle S\rangle^2}$ between the baryon number B and the strangeness S in a strongly interacting matter was first proposed in Refs.~\cite{Koch:2005vg,Haussler:2005ei,Cheng:2008zh} as a probe of the properties of the matter produced in relativistic heavy ion collisions. A later study suggested, however, that the strangeness population factor $\rm S_3 = $ \ssr measured in these collisions could serve as a better probe of the baryon number and strangeness correlation in the produced matter because of its different behaviors in QGP and the hadronic matter~\cite{Zhang:2009ba,Steinheimer:2012tb}. Experimentally, results on $\rm S_3$ show an increase of its value from heavy ion collisions at AGS~\cite{Armstrong:2002xh} to RHIC energy~\cite{Abelev:2010rv} and then a decrease to a small value in collisions at the LHC energy~\cite{Adam:2015yta}. Compared to the predictions from the statistical model~\cite{Andronic:2010qu,Cleymans:2011pe,Steinheimer:2012tb}, the values of $\rm S_3$ extracted from the RHIC data, which has large statistical uncertainties, are larger, and this has led to the questioning of the data and its interpretation. As to the different values of $\rm S_3$ measured at RHIC and LHC, a possible explanation was provided in Ref.~\cite{Sun:2015ulc} by assuming an early freeze-out of the $\Lambda$ than nucleons from the hadronic matter and a longer freeze-out time difference at RHIC than at LHC. This idea was further extended in Ref.~\cite{Sun:2018mqq} to study light nuclei production in relativistic heavy ion collisions by taking into consideration of their finite sizes compared to the size of produced hadronic matter at kinetic freeze out.

Since the first theoretical estimate on the abundance of hypernuclei that could be produced in heavy-ion collisions in 1970s~\cite{Kerman1973}, there have been many improved studies on this very interesting problem ~\cite{PhysRevC.32.326,Armstrong:2002xh,Abelev:2010rv,2013.NPA.913.170,Adam:2015yta,Liu:2017rjm,Zhang:2018euf}, see more details on recent topical review articles~\cite{2016.RMP.88.035004,CHEN20181,Braun-Munzinger:2018hat}. For the lightest hypernucleus $\rm ^3_{\Lambda}H$, the separation energy of its $\Lambda$ is very small in early measurements with a typical value of $130 \pm 50$ KeV~\cite{Davis:2005} but changes to a larger value of $410 \pm 120 \pm 110$ KeV in recent measurements using more precise method~\cite{Adam:2019phl}. Since this value is significantly smaller than that of normal nuclei with a similar mass number~\cite{Liu:2019mlm}, the \hyt can be considered as a loosely bound $\rm d-\Lambda$ 2-body system, and the ratio $\rm S_2=$\sr can also be used as an observable for probing the correlation between baryon and strangeness in relativistic heavy ion collisions.

In this paper, we study the ratios $\rm S_3=$\ssr and $\rm S_2=$\sr in the framework of the coalescence model. We first revisit the study of $\rm S_3$ and find that the discrepancy between the ALICE and STAR measurements may be partially due to the difference in the primordial proton yield used in the two analyses. We then show that the ratio $\rm S_2$, particularly its ratio $\rm S_2/B_2$ with respect to $\rm B_2$, which is the coalescence parameter for the production of a deuteron from a proton and neutron pair, is a cleaner probe of the baryon-strangeness correlation in the produced hadronic matter from relativistic heavy ion collisions.

\section{The $\rm S_3=$\ssr ratio in relativistic heavy ion collisions}

In this Section, we first review the experimental data on $\rm S_3$ from relativistic heavy ion collisions and
then use the coalescence model to extract from these data the correlation between $\Lambda$ and nucleon density fluctuations in the produced matter.

\subsection{Experimental results on $\rm S_3$}

For the value of $\rm S_3$, it was measured to be $0.36 \pm 0.26$ in central 11.5$A$ GeV/c Au + Pt collisions~\cite{Armstrong:2002xh} and increases to $1.08 \pm 0.22 \pm 0.16$ for Au+Au collisions at $\rm \sqrt{s_{NN}}$ = 200 GeV with a mixed event sample of central trigger and minimum bias trigger~\cite{Abelev:2010rv}. The measured value of $\rm S_3$ decreases, however, to $0.60 \pm 0.13 \pm 0.21$ in central Pb-Pb collisions at $\rm \sqrt{s_{NN}}$ = 2.76 TeV~\cite{Adam:2015yta}. Preliminary data with an improved precision from Au+Au collisions at $\rm \sqrt{s_{NN}}$ = 200 GeV results in a~20\% reduction of the value of $\rm S_3$~\cite{ZHU2013551c}, which makes the results from STAR and ALICE comparable within their experimental uncertainties. Replacing the proton yield in the $\rm S_3$ ratio in the STAR analysis, which is based on the subtraction of protons from measured hyperon data, with the proton yield in the PHENIX data, which is obtained from a theoretical model for the same collision system and energy~\cite{Adler:2003cb}, also reduces the value of $\rm S_3$ at $\rm \sqrt{s_{NN}}$ = 200 GeV to approximately same as the value measured by ALICE~\cite{Abelev:2010rv,ZHU2013551c,Adam:2015yta}. The difference between the results from STAR and ALICE is thus partially due to the different treatments in the subtraction of the weak decay contribution to the primordial proton yield. Table \ref{Tab1} summarizes the published $\rm S_3$ results, together with the value using the proton yield from the PHENIX data in Au+Au collisions at $\rm \sqrt{s_{NN}}$ = 200 GeV. It is seen that the values of $\rm S_3$ from STAR and ALICE are now comparable within their large uncertainties.

\begin{table*}[htbp]
\centering
\caption{Values of $\rm S_3$ from AGS, STAR and ALICE, where PH means that the proton yield is taken from PHENIX~\cite{Adler:2003cb}. See texts for details.}
\begin{tabular}{lccccccccc}
\hline
Experiment      &$\rm S_3$\\
\hline  \hline
AGS           &$0.36 \pm 0.26$    \\
STAR          &$1.08 \pm 0.22 \pm 0.16$    \\
STAR + PH &$0.90 \pm 0.22 \pm 0.15$   \\
ALICE       &$0.60  \pm 0.13 \pm 0.21$   \\
\hline
\end{tabular}
\label{Tab1}
\end{table*}

Experimentally, there also exists a puzzle related to the $\rm ^3_{\Lambda}H/\rm ^3He$ ratio~\cite{Cleymans:2011pe,Sun:2015ulc}. Its value is $0.82 \pm 0.16 \pm 0.12$ in $0-80\%$ centrality of Au+Au collisions at $\rm \sqrt{s_{NN}}$ = 200 GeV at RHIC~\cite{Abelev:2010rv}, which is considerably larger than the value of $0.47 \pm 0.10 \pm 0.13$ in Pb-Pb collisions at $\rm \sqrt{s_{NN}}$ = 2.76 TeV and 0-10\% centrality at the LHC~\cite{Adam:2015yta}. Although the preliminary measurement with improved precision by STAR in Au+Au collisions at $\rm \sqrt{s_{NN}}$ = 200 GeV results in a reduction of the $\rm ^3_{\Lambda}H/\rm ^3He$ ratio to a value comparable with that from ALICE, its large uncertainties~\cite{Adamczyk:2017buv} indicate that more precise measurements of \hyt in high energy heavy-ion collisions are needed.

\subsection{$\rm S_3$ in the coalescence model and the nucleon and $\Lambda$ density fluctuations}

To see the physics that can be extracted from the ratio $\rm S_3$, we adopt the coalescence model for the present study. According to the COAL-SH formula in Ref.~\cite{Sun:2018jhg}, the yield of certain nucleus consisting of $N_i$ number of the constituent species $i$ (proton, neutron and $\Lambda$) of mass $m_i$ from the kinetically freezed-out hadronic matter of local temperature T and volume $\rm V_{K}$ in a heavy ion collision can be written as
\begin{eqnarray}\label{yield}
\rm N_A = g_{rel}g_{size}g_A\left(\sum_i^A m_i\right)^{3/2}\bigg[\prod_{i=1}^A\frac{N_i}{m_i^{3/2}}\bigg]\times\prod_{i=1}^{A-1}\frac{(4\pi/\omega)^{3/2}}{V_{K}x(1+x^2)}\bigg(\frac{x^2}{1+x^2}\bigg)^{l_i}G(l_i,x).
\end{eqnarray}
In the above, $\rm g_{A}=(2S+1)/(\prod_{i=1}^A(2s_i+1))$ is the coalescence factor for $A$ nucleons and/or $\Lambda$ of spin $s_i=1/2$ to form a nucleus of spin S, $\rm g_{rel}$ is the relativistic correction to the effective volume in momentum space and is set to 1 in the present study due to the much larger nucleon mass than the effective temperature of the hadronic matter at kinetic freeze out, $\rm g_{size}$ is the correction due to the finite size of produced nucleus and is also taken to be 1 in our study because of the much larger size of the hadronic matter than the sizes of produced light nuclei. The symbols $l_i$ and $\omega$ denote, respectively, the orbital angular momentum of the nucleon or $\Lambda$ in the nucleus and the oscillator constant used in its wave function. For the value of $\rm x=(2T/\omega)^{1/2}$, it is significantly larger than one because of the much larger size of the nucleus than the thermal wavelength of its constituents in the hadronic matter. Since the light nuclei considered in the present study all involve only the $l=0$ $s$-wave, the suppression factor $G(l_i,x)$ due to the orbital angular momentum in the above equation is simply one.

For the production of \hyt and $\rm ^3He$, their yields according to Eq.(\ref{yield}) after taking into account the approximations mentioned above are given by
\begin{eqnarray}\label{3}
&&\rm N_{^3_{\Lambda}H} = g_{^3_{\Lambda}H}\frac{(m_{\Lambda}+m_p+m_n)^{3/2}}{m_{\Lambda}^{3/2}m_p^{3/2}m_n^{3/2}}\left(\frac{2\pi}{T}\right)^3\frac{N_{\Lambda}N_pN_n}{V_{K}^2},\notag\\
&&\rm N_{^3He} = g_{^3He}\frac{(2m_p+m_n)^{3/2}}{m_p^{3}m_n^{3/2}}\left(\frac{2\pi}{T}\right)^3\frac{N_p^2N_n}{V_{K}^2}.
\end{eqnarray}
Allowing possible nucleon and $\Lambda$ density fluctuations in heavy ion collisions at lower energies due to the spinodal instability during the QGP to hadronic matter phase transition~\cite{Sun:2018jhg,Ko:2016ioz,Shao:2019xpj,Jin:2019}, we rewrite the density distributions of nucleons and $\Lambda$ as
\begin{eqnarray}
&& n(\vec{r}) = \frac{1}{V_K} \int n(\vec{r})d\vec{r} + \delta n(\vec{r})= \langle n\rangle + \delta n(\vec{r}).
\end{eqnarray}
According to Ref.~\cite{Sun:2017xrx}, this modifies Eq.(\ref{3}) to
\begin{eqnarray}
&&\rm N_{^3_{\Lambda}H} = g_{^3_{\Lambda}H}\frac{(m_{\Lambda}+m_p+m_n)^{3/2}}{m_{\Lambda}^{3/2}m_p^{3/2}m_n^{3/2}}\bigg(\frac{2\pi}{T}\bigg)^{3}\frac{N_{\Lambda}N_{p}N_n}{V_{K}^2}(1+\alpha_{\Lambda p}+\alpha_{\Lambda n}+\alpha_{np}),\notag\\
&&\notag\\
&&\rm N_{^3He} = g_{^3He}\frac{(2m_p+m_n)^{3/2}}{m_p^{3}m_n^{3/2}}\bigg(\frac{2\pi}{T}\bigg)^{3}\frac{N_p^2N_n}{V_{K}^2}(1+\Delta p+2\alpha_{np}),
\label{he3}
\end{eqnarray}
if one neglects higher-order correlation coefficients of density fluctuations. In the above, $\rm \Delta p = \langle (\delta p)^2\rangle/\langle p\rangle^2$ is the proton relative density fluctuation, and $\rm \alpha_{\Lambda p}$, $\rm \alpha_{\Lambda n}$, and $\rm \alpha_{np}$ are, respectively, the $\Lambda$-proton, $\Lambda$-neutron, and proton-neutron density fluctuation correlation coefficients $\rm \alpha_{n_1 n_2} = \langle \delta n_1 \delta n_2\rangle/(\langle n_1\rangle\langle n_2\rangle)$ with $n_1$ and $n_2$ denoting $\Lambda$ or nucleon.

Taking the same mass for proton and neutron, i.e., $m_p=m_n=m$, the yield ratio $\rm S_3 = $ \ssr is then
\begin{eqnarray}
\rm S_3 = g \frac{1+\alpha_{\Lambda p}+\alpha_{\Lambda n}+\alpha_{np}}{1+\Delta p+2\alpha_{np}},
\label{s3-eq}
\end{eqnarray}
with $\rm g=\left(\frac{m_\Lambda+2m}{3m_{\Lambda}}\right)^{3/2}\approx 0.845$. From the above equation, the sum of the correlation coefficients $\alpha_{\Lambda p}$+$\alpha_{\Lambda n}$ between the $\Lambda$ density and the proton or neutron density fluctuations can be determined from $\rm S_3$, $\rm \Delta p$, and $\rm \alpha_{np}$ according to
\begin{eqnarray}
\rm \alpha_{\Lambda p} + \alpha_{\Lambda n} = \frac{S_3}{g} \times (1+\Delta p+2\alpha_{np}) - \alpha_{np} - 1.
\label{lp+ln}
\end{eqnarray}

For the value of $\rm \alpha_{np}$, we follow the method in Ref.~\cite{Sun:2018jhg} by using the deuteron yield after including in the coalescence formula COAL-SH the proton and neutron density fluctuations, i.e.,
\begin{eqnarray}
\rm N_d = 2^{3/2}g_d\bigg(\frac{2\pi}{mT}\bigg)^{3/2}\frac{N_pN_n}{V_{K}}(1+\alpha_{np}).
\end{eqnarray}
In terms of $\rm g_{d-p} =\frac{1}{2^{3/2}g_d(2\pi)^3}=\frac{2^{1/2}}{3(2\pi)^3} \approx 0.0019$, $\rm \mathcal{O}_{d-p} = N_d/N_p^2$, $\rm R_{np} = N_p/N_n$, and $\rm V_{ph} = (2\pi mT)^{3/2}V_{K}$, the value of $\rm \alpha_{np}$ can then be calculated from
\begin{eqnarray}
\rm \alpha_{np} = g_{d-p}R_{np}V_{ph}\mathcal{O}_{d-p} - 1.
\end{eqnarray}

For the proton density fluctuation $\Delta p$, we consider the ratio
\begin{eqnarray}
\rm \frac{N_{^3He}}{N_p^3} \times \frac{N_p}{N_n} = 3^{3/2}g_{^3He}\left(\frac{2\pi}{mT}\right)^3\frac{1}{V_{K}^2}(1+\Delta p+2\alpha_{np}),
\end{eqnarray}
and determine it from the relation
\begin{eqnarray}
\rm \Delta p = g_{^3He-p}V_{ph}^2R_{np}\mathcal{O}_{^3He-p} - 2\alpha_{np} - 1,
\end{eqnarray}
where $\rm g_{^3He-p} = \frac{1}{3^{3/2}g_{^3{\rm He}}(2\pi)^6}=\frac{4}{3^{3/2}(2\pi)^6} \approx 1.25\times10^{-5}$ and $\rm \mathcal{O}_{^3He-p} = N_{^3He}/N_p^3$, and the factors $\rm V_{ph}$, $\rm R_{np}$ and $\rm \alpha_{np}$ are the same as above.

\begin{table*}[htbp]
\centering
\caption{Values of parameters used for 0-10\% central collisions.}
\begin{tabular}{lccccccccc}
\hline
 $\rm \sqrt{s_{NN}} (GeV)$~~      &~~$\rm T_{ch} (GeV)$~~   &~~$\rm V_{ch} (fm^3)$~~  &~~$\rm R_{np}$~~   &~~$\rm \alpha_{np}$~~  &~~~~~$\lambda$~~\\ 
\hline  \hline
4.9            &0.132    &640     &0.925  &$-0.781\pm0.026$  &~~~~~2.23\\
7.7            &0.144    &806     &0.966  &$-0.744\pm0.024$   &~~~~~2.60\\ 
11.5           &0.151    &875     &0.977  &$-0.763\pm0.019$   &~~~~~2.76 \\ 
19.6           &0.158    &843     &0.987  &$-0.830\pm0.014$   &~~~~~2.92 \\ 
27             &0.160    &846     &0.988  &$-0.848\pm0.012$   &~~~~~2.97 \\ 
39             &0.160    &951    &0.990 &$-0.834\pm0.013$     &~~~~~3.00\\ 
62.4           &0.164    &1215    &0.992 &$-0.792\pm0.037$    &~~~~~3.16 \\
200            &0.168    &1334    &0.992  &$-0.726\pm0.038$   &~~~~~3.30 \\ 
2760           &0.156    &4320    &1.00  &$-0.717\pm0.023$   &~~~~~2.94\\ 
\hline
\end{tabular}
\label{Tab2}
\end{table*}

For the effective phase-space volume $\rm V_{ph}$ occupied by nucleons in the hadronic matter at kinetic 
freeze-out, they can be evaluated from its value at chemical freeze-out by using the relation $\rm T^{3/2}V_{K} 
= \lambda T_{ch}^{3/2}V_{ch}$, where $\rm T_{\rm ch}$ and $\rm V_{\rm ch}$ are, respectively, the 
temperature and volume of the system at chemical freeze-out, and $\lambda$ is a parameter. For collisions at RHIC energies, we take the value of $\rm T_{ch}$ from the grand canonical ensemble fits to the particle yields in Ref.~\cite{Adamczyk:2017iwn} and that of $\rm V_{ch}$ to be $\rm V_{ch} = 4 \mathrm \pi R^3/3$ as in Ref.~\cite{Adamczyk:2017iwn} for collisions at various centralities except for collisions at $0-80\%$ centrality, where it is taken to be proportional to the charged particle multiplicity obtained from Ref.~\cite{Abelev:2008ab}. Using the results from Ref.~\cite{Adamczyk:2017iwn} based on the strangeness suppressed canonical ensemble fits gives almost the same results. The values of $\rm T_{ch}$ and $\rm V_{ch}$ used in the present study for collision at AGS energy are taken from Ref.~\cite{Andronic:2005yp}, and for collision at the LHC energy, they are taken from the COAL-SH model used in Ref.~\cite{Sun:2017ooe}. In our calculations, all hadrons are taken as point particles. Including an exclusive volume for each hadron would give a larger value for $\rm V_{ch} $~\cite{Andronic:2005yp}. For the value of $\lambda$, it is determined by assuming that the entropy per baryon is the same at the chemical and the kinetic freeze out after including in the hadronic matter all the resonances in PDG. This approach thus differs from the naive approach of a hadronic matter of constant number of nucleons expanding isentropically after chemical freeze-out, which would give a $\lambda$ of value of unity, and also that in Ref.~\cite{Sun:2018jhg} based on a multiphase transport model.  As to the value of $\rm R_{np} = N_p/N_n$, it can be determined from measured ratio of charged pions according to the relation $\rm N_p/N_n = (N_{\pi^+}/N_{\pi^-})^{1/2}$ from the statistical model. In Table \ref{Tab2}, we summarize the values of above parameters used in the present study for central heavy ion collisions. For reference, we also provide in Table~\ref{Tab22} their values for collisions at the centralities of 0-80\% or 0-60\%. 

\begin{table*}[htbp]
\centering
\caption{Same as Table~\ref{Tab2} for values of parameters used in calculations for 0-80\% or 0-60\% centralities in collisions at RHIC energies and the LHC energy, respectively.}
\begin{tabular}{lccccccccc}
\hline
 $\rm \sqrt{s_{NN}} (GeV)$~~      &~~$\rm T_{ch} (GeV)$~~   &~~$\rm V_{ch} (fm^3)$~~  &~~$\rm R_{np}$~~   &~~$\rm \alpha_{np}$~~ &~~~~~$\lambda$~~\\ 
\hline  \hline
7.7            &0.144    &268     &0.966  &$-0.775\pm0.020$    &~~~~~2.60\\ 
11.5           &0.151    &292     &0.977  &$-0.793\pm0.018$    &~~~~~2.76\\ 
19.6           &0.158    &281     &0.987  &$-0.851\pm0.014$    &~~~~~2.92\\ 
27             &0.160    &282     &0.988  &$-0.867\pm0.012$    &~~~~~2.97\\ 
39             &0.160    &317     &0.990  &$-0.859\pm0.013$    &~~~~~3.00\\ 
200            &0.168    &445     &0.992  &$-0.747\pm0.036$    &~~~~~3.30\\ 
2760         &0.156      &1800     &1.00  &$-0.710\pm0.024$    &~~~~~2.94\\ 
\hline
\end{tabular}
\label{Tab22}
\end{table*}

As shown in Tables \ref{Tab2} and \ref{Tab22}, the extracted values for the neutron and proton density fluctuation correlation $\alpha_{np}$ are negative with appreciable magnitude, indicating that neutrons and protons are anti-correlated in the matter produced in relativistic heavy ion collisions, similar to that found in Ref.~\cite{Sun:2018jhg}. For the proton density fluctuation $\rm \Delta p$, the extracted values are 0.656 $\pm$ 0.049, 0.536 $\pm$ 0.083, and 0.727 $\pm$ 0.085 for collisions at $\rm \sqrt{s_{NN}}$ = 4.9, 200, and 2760 GeV, respectively~\cite{Armstrong:2002xh,Abelev:2010rv,Adam:2015vda}. It shows a non-monotonic behavior as a function of collision energy from $\rm \sqrt{s_{NN}}$ = 7.7 GeV to 200 GeV using the preliminary data of $\rm ^3He$ yield from Ref.~\cite{ZHU2013551c}, similar to that of the neutron density fluctuation extracted from the yield ratio of $\rm \frac{N_t N_p}{N_{d}^2}$~\cite{Zhang:2020ewj}. From measured values of $\rm S_3$ and extracted values for $\rm \alpha_{np}$ and $\rm \Delta p$, one can then determine the values of $\rm \alpha_{\Lambda n}+\alpha_{\Lambda p}$ from heavy ion collisions at various energies according to Eq.(\ref{lp+ln}). These results will be shown in the next Section to compare with the correlation coefficient $\rm \alpha_{\Lambda d}$ of the $\Lambda$ and deuteron density fluctuations that is extracted from measured $\rm S_2=$\sr ratio.

\section{The $\rm S_2=$ \sr ratio}

In this Section, we consider the $\rm S_2=$ \sr ratio in the framework of the coalescence model and its ratio $\rm S_2/B_2$ with respect to the coalescence parameter $\rm B_2$ for the production of deuteron from the coalescence of proton and neutron as well as the extracted correlation coefficient $\rm \alpha_{\Lambda d}$ between the $\Lambda$ and deuteron density fluctuations using the experimental data from RHIC and the LHC.

\subsection{The $\rm S_2=$ \sr ratio in the coalescence model}

Approximating \hyt as a bound system of $\Lambda$ and deuteron, the yield of \hyt in heavy ion collisions can be calculated from the coalescence of $\Lambda$ and deuteron using Eq.(\ref{yield}). Including the effect of deuteron and $\Lambda$ density fluctuations, it is given by
\begin{eqnarray}
\rm N_{^3_{\Lambda}H} = g_{^3_{\Lambda}H}\frac{(m_{\Lambda}+m_d)^{3/2}}{m_{\Lambda}^{3/2}m_d^{3/2}}\bigg(\frac{2\pi}{T}\bigg)^{3/2}\frac{N_{\Lambda}N_d}{V{_K}}(1+\alpha_{\Lambda d}),
\label{h3l}
\end{eqnarray}
with $\rm \alpha_{\Lambda d} = \langle \delta n_{\Lambda}\delta n_{d}\rangle/(\langle n_{\Lambda}\rangle\langle n_{d}\rangle)$ being the correlation coefficient between deuteron and $\Lambda$ density fluctuations. The $\rm S_2$ ratio is then
\begin{eqnarray}\label{s2-eq}
\rm S_2=\rm \frac{N_{^3_{\Lambda}H}}{N_\Lambda N_d} = g_{^3_{\Lambda}H}\frac{(m_{\Lambda}+m_d)^{3/2}}{m_{\Lambda}^{3/2}m_d^{3/2}}\bigg(\frac{2\pi}{T}\bigg)^{3/2}\frac{1}{V{_K}}(1+\alpha_{\Lambda d}),
\end{eqnarray}
with $\rm g_{S_2}= \left(\frac{1}{3}\frac{(m_{\Lambda}+m_d)^{3/2}}{m_{\Lambda}^{3/2}m_d^{3/2}}(2\pi)^{3/2}\right)^{-1} \approx 0.12$, and from which we can express the density fluctuation correlation coefficient $\rm \alpha_{\Lambda d}$ in terms of $\rm S_2$ as
\begin{eqnarray}
\rm \alpha_{\Lambda d} = g_{S_2} S_2  T^{3/2}V_{K} - 1 .
\label{lad1}
\end{eqnarray}

\begin{figure}[!htb]
\centering
\includegraphics[scale=0.4]{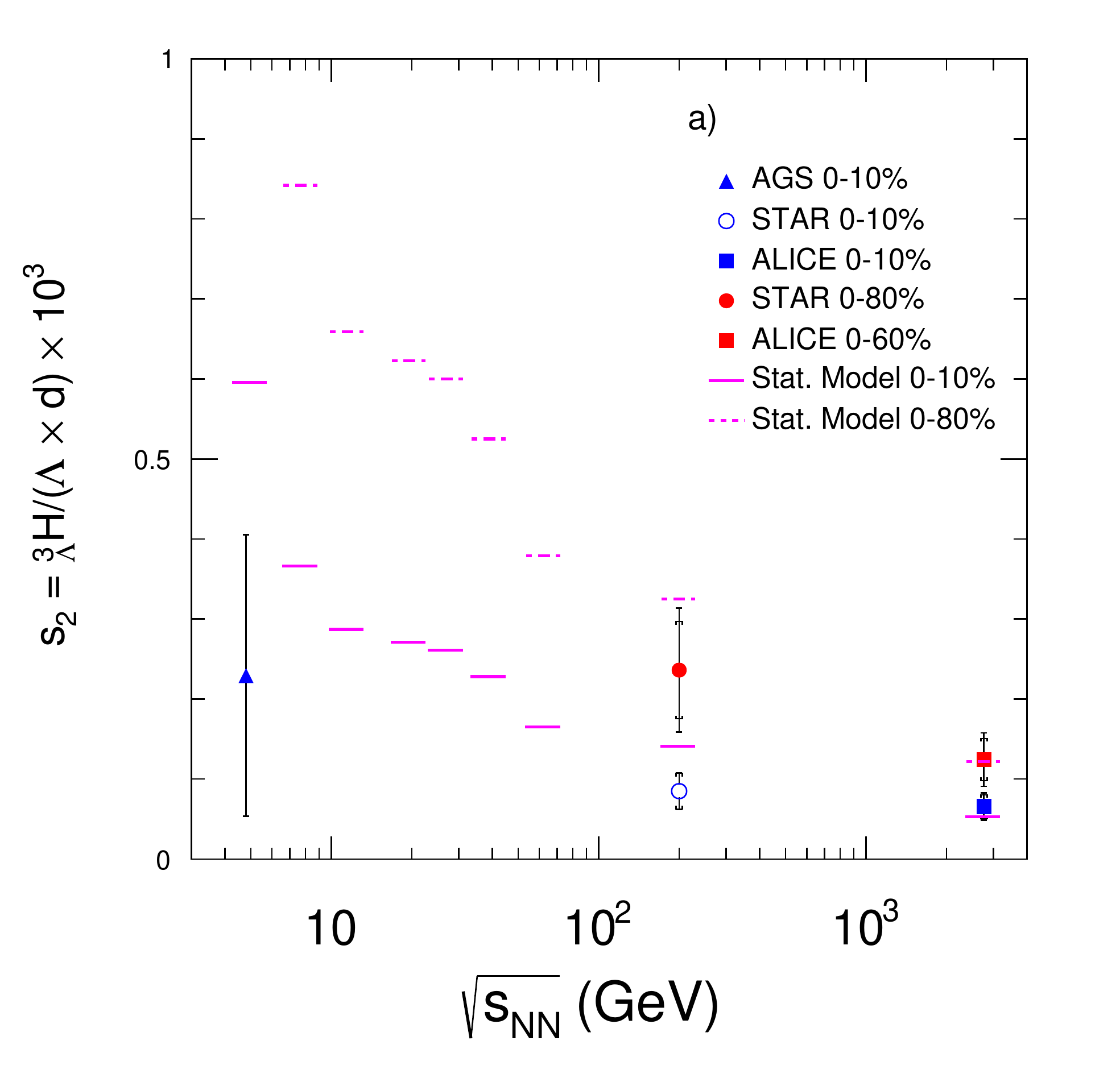}
\includegraphics[scale=0.4]{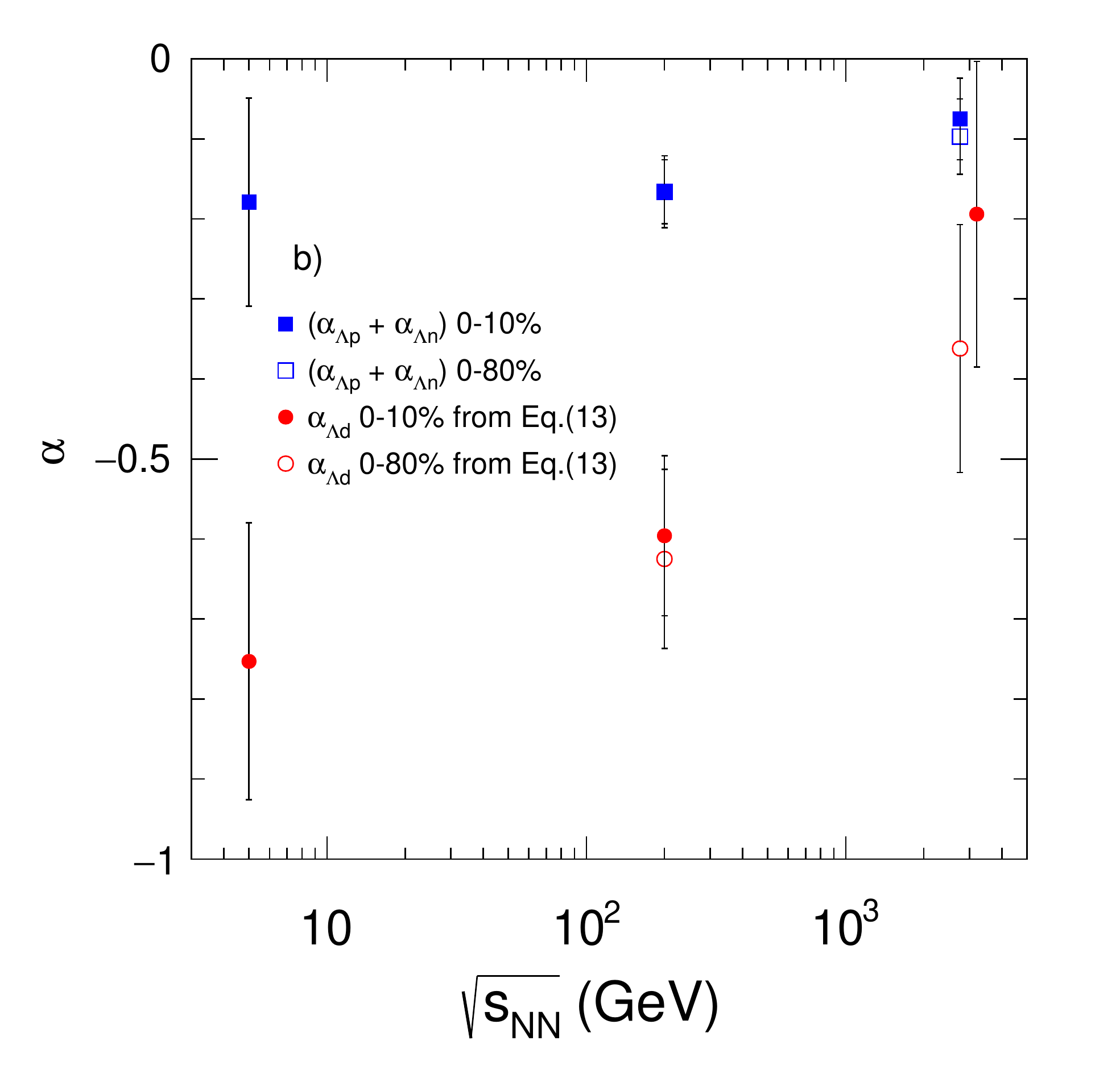}
\caption{Collision energy dependence of the $\rm S_2=$\sr ratio (left window) and the density fluctuation correlation coefficients $\rm \alpha_{\Lambda p} + \alpha_{\Lambda n}$ and $\rm \alpha_{\Lambda d}$ (right window) extracted from the experimental data (symbols) and calculated from the statistical model (horizontal bars) using parameters in Tables~\ref{Tab2} and \ref{Tab22}. See texts for details.}
\label{s2}
\end{figure}

In the left window of Fig.~\ref{s2}, we show by solid symbols the extracted ratio $\rm S_2=$\sr using experimental data from AGS~\cite{Armstrong:2002xh,Armstrong:2000gz,Albergo:2002tn}, RHIC~\cite{Abelev:2010rv,Adam:2019wnb,Agakishiev:2011ar} and LHC~\cite{Adam:2015yta,Adam:2015vda,Abelev:2013xaa}. Open symbol represents the result for the \hyt yield in collision at 0-10\% centrality where the RHIC data is obtained from multiplying the measured \hyt data at 0-80\% centrality by a factor of 3. We have checked using data available from the RHIC BES program that the yield ratio of deuteron to $\Lambda$ is a factor 3 larger in collisions at 0-10\% centrality than at 0-80\% centrality, independent of the collision energy~\cite{Adam:2019wnb,Adam:2019koz}. Solid and dashed lines in the left window of Fig.~\ref{s2} are results from statistical model calculations~\cite{Sun:2018jhg} using the parameters in Tables~\ref{Tab2} and \ref{Tab22}. One sees that although there are missing data points in the large collision energy range, the ratio $\rm S_2$ in collisions at 0-10\% centrality seems independent of the collision energy $\rm \sqrt{s_{NN}}$ considering the large uncertainty of the AGS data. Also, the model calculation describes reasonably well the data at the LHC energy but over-predicts the results for collisions at AGS and RHIC energies.

Right window of Fig.~\ref{s2} shows the value of the correlation coefficients $\rm \alpha_{\Lambda d}$ as a function of the collision energy for collisions at $0-10\%$ centrality (solid symbols) and at $0-80\%$ centrality (open symbols). The large uncertainty at $\rm \sqrt{s_{NN}}$ = 2.76 TeV is due to the standard error propagation from the large volume $\rm V_{ch}$ used in collisions at the LHC energy. Within current experimental uncertainties, the value of $\rm \alpha_{\Lambda d}$, which is negative and thus indicates an anti-correlation between the $\Lambda$ and deuteron density fluctuations, becomes slightly less negative as the collision energy increases and approaches towards zero at $\rm \sqrt{s_{NN}}$ = 2.76 TeV. The negative $\rm \alpha_{\Lambda d}$ and the negative $\rm \alpha_{np}$\, shown in Tables \ref{Tab2} and \ref{Tab22} could be due to an underestimate of the value of the $\lambda$ parameter or the kinetic freeze-out volume used in our study. To fully understand these results requires detailed studies based on microscopic models for light cluster production in high energy heavy-ion collisions~\cite{Oh:2007vf,Oh:2009gx,Oliinychenko:2018ugs}. Compared with the correlation coefficient $\rm \alpha_{\Lambda p}+\alpha_{\Lambda n}$ extracted from $\rm S_3$ shown by solid and open squares for collisions at centralities of 0-10\% and 0-80\%, respectively, which seems to vary very little over a broad range of collision energies, the deviation of its value from zero is larger than that of $\rm \alpha_{\Lambda p}+\alpha_{\Lambda n}$, as it shows a more visible $\rm \sqrt{s_{NN}}$ dependence. This may suggest that $\rm \alpha_{\Lambda d}$ is a cleaner observable than $\rm \alpha_{\Lambda p}+\alpha_{\Lambda n}$ for studying the $\rm \sqrt{s_{NN}}$ dependence of baryon density fluctuations and their correlations as seen from the comparison of Eq.~(\ref{s2-eq}) to Eq.~(\ref{s3-eq}). Future experimental measurements in a broad collision energy range from AGS to RHIC will be very useful for shedding lights on the underlying physics.

\subsection{The $\rm S_2/B_2$ ratio}

In the coalescence model, the yield ratio $\rm S_2 = N_{^3_{\Lambda}H}/(N_{\Lambda}N_d)$ is the coalescence parameter for the production of $\rm ^3_{\Lambda}H$ if it is considered as a bound system of $\Lambda$ and deuteron. Because of the strangeness carried by $\Lambda$, the $\rm S_2$ may be different from the coalescence parameter $\rm B_2$ for the production of deuteron from the coalescence of proton and neutron~\cite{Csernai:1986qf,CHEN20181,Dong2018,Zhao:2018lyf}.

From Eq.(\ref{s2-eq}) for $\rm S_2$ and a similar equation for $\rm B_2$, given by
\begin{eqnarray}
\rm B_2=\rm \frac{N_d}{N_p N_n} = g_{d}\frac{1}{m_p^{3/2}}\bigg(\frac{2\pi}{T}\bigg)^{3/2}\frac{1}{V_{K}}(1+\alpha_{np}),
\end{eqnarray}
taking $\rm N_n=N_p$ then leads to
\begin{eqnarray}
\rm \frac{S_2}{B_2} = \frac{N_{^3_{\Lambda}H}}{N_\Lambda N_d}\bigg/\frac{N_d}{N_p N_n} = g\frac{1+\alpha_{\Lambda d}}{1+\alpha_{np}},
\label{eq2-sb2}
\end{eqnarray}
where $\rm g = \frac{g_{^3_{\Lambda}H}}{g_d}\frac{m_p^{3/2}(m_{\Lambda}+m_d)^{3/2}}{m_{\Lambda}^{3/2}m_d^{3/2}} \approx 0.23$. The ratio $\rm S_2/B_2$ thus carries information on the difference between $\rm \alpha_{np}$ and $\rm \alpha_{\Lambda d}$ and thus the difference between the baryon-baryon correlation and the baryon-strangeness correlation.

Similarly, we can introduce the coalescence parameter $\rm B_3=\frac{N_{^3He}}{N_pN_pN_n}$ for the production of \he from the three-body coalescence of two protons and a neutron, and the coalescence parameter $\rm B_{s3}=\frac{N_{^3_{\Lambda}H}}{N_{\Lambda}N_pN_n}$ for the production of \hyt from the three-body coalescence of $\Lambda$, proton and neutron. Their ratio is exactly the $\rm S_3$ discussed in Section II.  Since the ratio $\rm S_2/B_2$ does not involve the proton density fluctuation $\rm \Delta p$ and other mixed density fluctuation correlations, it seems a more sensitive observable than $\rm S_3$ for studying the $\Lambda$ density fluctuation.

\begin{figure}[!htb]
\centering
\includegraphics[scale=0.4]{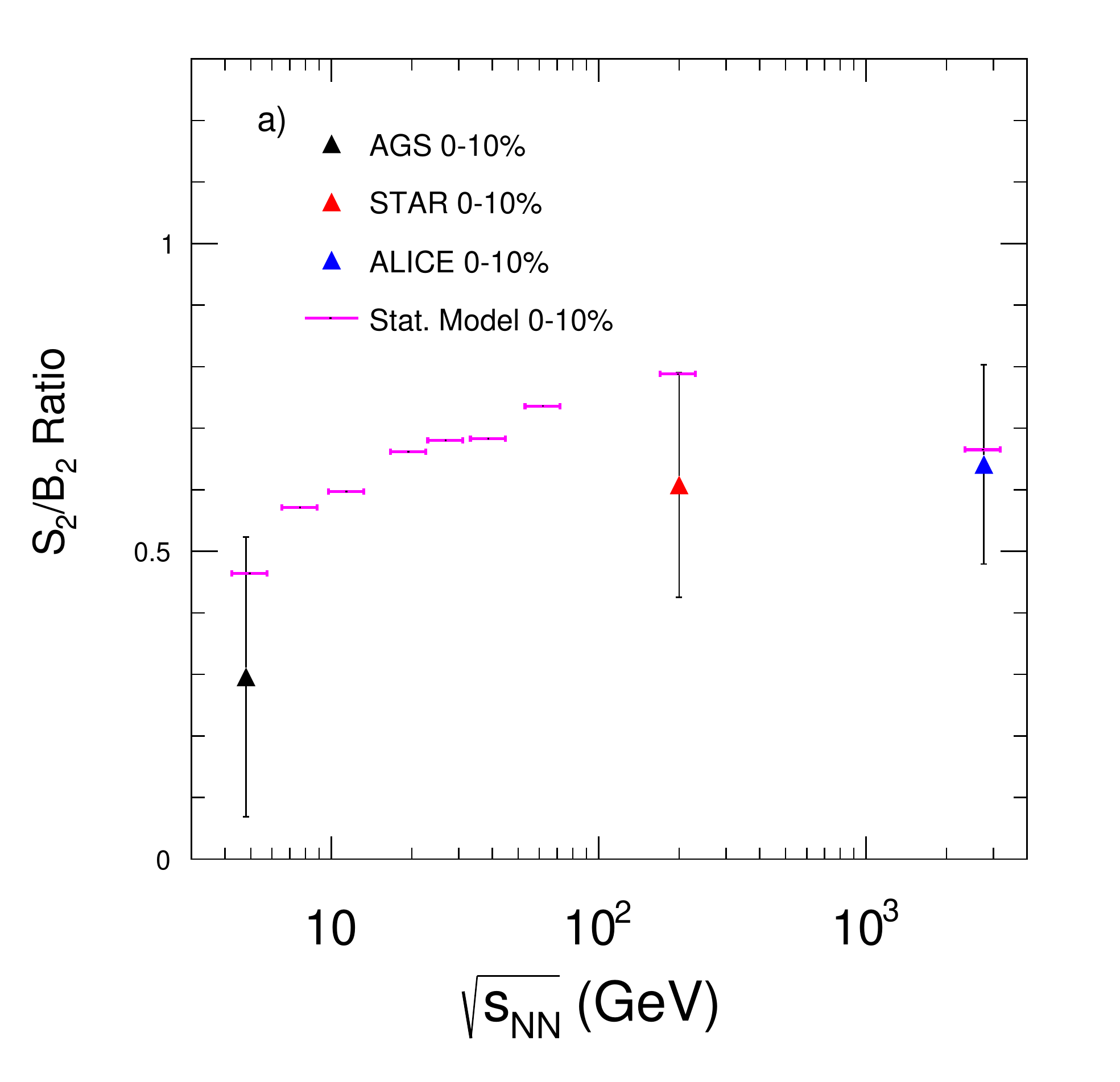}
\includegraphics[scale=0.4]{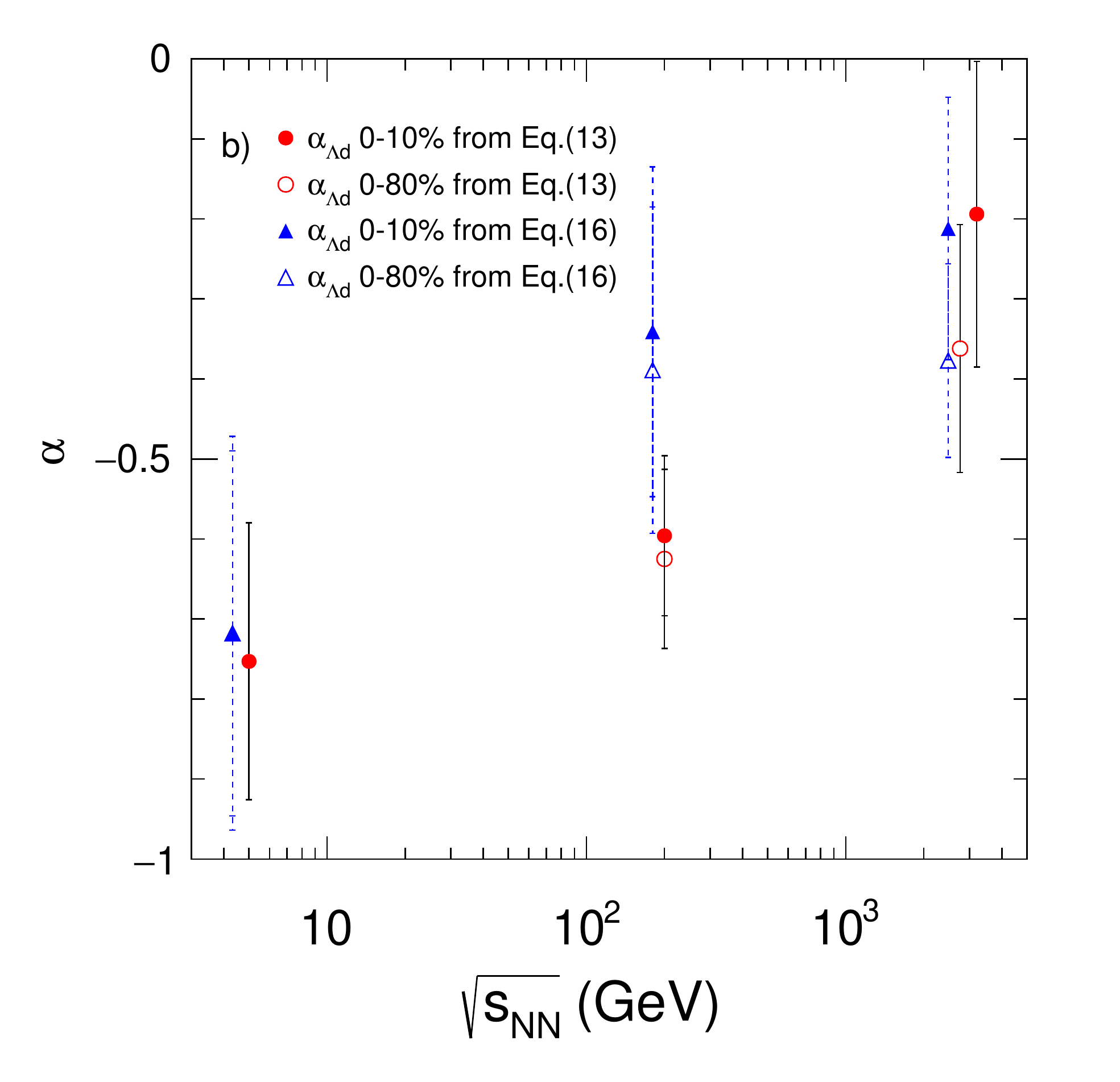}
\caption{Left window: The $\rm S_2/B_2$ ratio extracted from experimental data (solid triangles) and predicted by the statistical model (solid horizontal bars) for collision at $0-10\%$ centrality~\cite{Adam:2019wnb,Adam:2015vda}. Right window: Values of $\alpha_{\Lambda d}$ extracted from experimental results according to Eqs.(\ref{lad2}) and (\ref{lad1}).}
\label{sb2}
\end{figure}

Left window of Fig.~\ref{sb2} shows results for the yield ratio $\rm S_2/B_2$ = $\rm N_{^3_{\Lambda}H}/(N_{\Lambda}N_d)/(N_d/N_p^2)$ from experimental data (solid triangles) and predicted by the statistical model (solid horizontal bars) using the proton, deuteron and $\rm ^3_{\Lambda}H$ yields from the full $\rm p_T$ range. It is seen that the measured yield ratio increases slightly with increasing collision energy as in predictions from the statistical model. We note that the value of $\rm B_2$ has also been determined in experiments from the proton, neutron and deuteron momentum spectra in a small $\rm p_T$ window~\cite{Adam:2019wnb}. The $\rm S_2/B_2$ ratio obtained from collisions at the LHC energy for momentum per constituent $\rm p_{T}/A = 1.4~GeV/c$~\cite{Adam:2015yta,Adam:2015vda,Abelev:2013xaa} is $0.899\pm0.171$. Within their uncertainties, this value is similar to that obtained using yields from the full $\rm p_T$ range. However, the $\rm S_2/B_2$ ratio measured for particular $\rm p_T$ bins are unavailable from experiments in the energy range between those available at AGS and RHIC, and this is due to the lack of \hyt $\rm p_T$ spectra at these energies. Future measurements of \hyt spectra over a broad energy range are needed for extracting the $\Lambda$ and deuteron density correlation coefficient $\rm \alpha_{\Lambda d}$ discussed below.

The $\Lambda$ and deuteron correlation coefficient $\rm \alpha_{\Lambda d}$ can also be extracted from the yield ratio $\rm S_2/B_2$ given in Eq.~(\ref{eq2-sb2}), that is
\begin{eqnarray}
\rm \alpha_{\Lambda d} = \bigg(\frac{N_{^3_{\Lambda}H}}{N_{\Lambda}N_d(N_d/N_p^2)}\bigg/g\bigg) \times (1+\alpha_{np}) - 1,
\label{lad2}
\end{eqnarray}
by taking advantage of the empirical fact that the neutron and proton density fluctuation correlation $\rm \alpha_{np}$ is less affected by $\rm T$ and $\rm V_{K}$. Shown in the right window of Fig. \ref{sb2} by triangles are the values of $\alpha_{\Lambda d}$ extracted from the experimental results using Eq.(\ref{lad2}). They are seen to have similar values as those obtained from Eq.(\ref{lad1}) using $\rm S_2 = \frac{N_{^3_\Lambda H}}{N_\Lambda N_d}$, which are shown by filled circles and also in Fig.~\ref{s2} where it is compared with the density fluctuation correlation coefficient $\rm \alpha_{\Lambda p} + \alpha_{\Lambda n}$ extracted from $\rm S_3 = \frac{N_{^3_\Lambda H}/N_\Lambda}{N_{^3 He}/N_p}$.

\section{Conclusion}

In summary, we have argued that both the ratio $\rm S_2$ and the ratio $\rm S_2/B_2$, where $\rm S_2$ and $\rm B_2$ are, respectively, the coalescence parameter for the production of hypertriton from $\Lambda$ and deuteron and of deuteron from proton and neutron, are more sensitive observables than the previously proposed ratio $\rm S_3$=\ssr for studying the local baryon-strangeness correlation in the matter produced in relativistic heavy ion collisions. We have substantiated this argument in the framework of baryon coalescence by demonstrating that the correlation coefficient $\rm \alpha_{\Lambda d}$ between $\Lambda$ and deuteron density fluctuations extracted from measured $\rm S_2/B_2$ shows a stronger dependence on the energy of heavy ion collisions than the correlation coefficients $\rm \alpha_{\Lambda p} + \alpha_{\Lambda n}$ between $\Lambda$ and nucleon density fluctuations extracted from the measured $\rm S_3$. Experimental measurement of the ratio $\rm S_2/B_2$ is expected to provide a promising way to study the strangeness and baryon correlation in the matter produced from heavy ion collisions as the collision energy or the baryon chemical potential of produced matter is varied, which in turn can shed lights on the properties of the QGP to hadronic matter phase transition during the collisions.
\section{Acknowledgement}

The work of T.S. and J.C. was supported in part by the National Natural Science Foundation of China under Contract Nos. 11890710, 11775288, 11421505 and 11520101004, while that of C.M.K. and K.J.S. was supported by the US Department of Energy under Contract No. DE-SC0015266 and the Welch Foundation under Grant No. A-1358.

\bibliography{myref}

\begin{thebibliography}{49}%
\makeatletter
\providecommand \@ifxundefined [1]{%
 \@ifx{#1\undefined}
}%
\providecommand \@ifnum [1]{%
 \ifnum #1\expandafter \@firstoftwo
 \else \expandafter \@secondoftwo
 \fi
}%
\providecommand \@ifx [1]{%
 \ifx #1\expandafter \@firstoftwo
 \else \expandafter \@secondoftwo
 \fi
}%
\providecommand \natexlab [1]{#1}%
\providecommand \enquote  [1]{``#1''}%
\providecommand \bibnamefont  [1]{#1}%
\providecommand \bibfnamefont [1]{#1}%
\providecommand \citenamefont [1]{#1}%
\providecommand \href@noop [0]{\@secondoftwo}%
\providecommand \href [0]{\begingroup \@sanitize@url \@href}%
\providecommand \@href[1]{\@@startlink{#1}\@@href}%
\providecommand \@@href[1]{\endgroup#1\@@endlink}%
\providecommand \@sanitize@url [0]{\catcode `\\12\catcode `\$12\catcode
  `\&12\catcode `\#12\catcode `\^12\catcode `\_12\catcode `\%12\relax}%
\providecommand \@@startlink[1]{}%
\providecommand \@@endlink[0]{}%
\providecommand \url  [0]{\begingroup\@sanitize@url \@url }%
\providecommand \@url [1]{\endgroup\@href {#1}{\urlprefix }}%
\providecommand \urlprefix  [0]{URL }%
\providecommand \Eprint [0]{\href }%
\providecommand \doibase [0]{http://dx.doi.org/}%
\providecommand \selectlanguage [0]{\@gobble}%
\providecommand \bibinfo  [0]{\@secondoftwo}%
\providecommand \bibfield  [0]{\@secondoftwo}%
\providecommand \translation [1]{[#1]}%
\providecommand \BibitemOpen [0]{}%
\providecommand \bibitemStop [0]{}%
\providecommand \bibitemNoStop [0]{.\EOS\space}%
\providecommand \EOS [0]{\spacefactor3000\relax}%
\providecommand \BibitemShut  [1]{\csname bibitem#1\endcsname}%
\let\auto@bib@innerbib\@empty
\bibitem [{\citenamefont {Koch}\ \emph {et~al.}(2005)\citenamefont {Koch},
  \citenamefont {Majumder},\ and\ \citenamefont {Randrup}}]{Koch:2005vg}%
  \BibitemOpen
  \bibfield  {author} {\bibinfo {author} {\bibfnamefont {V.}~\bibnamefont
  {Koch}}, \bibinfo {author} {\bibfnamefont {A.}~\bibnamefont {Majumder}}, \
  and\ \bibinfo {author} {\bibfnamefont {J.}~\bibnamefont {Randrup}},\ }\href
  {\doibase 10.1103/PhysRevLett.95.182301} {\bibfield  {journal} {\bibinfo
  {journal} {Phys. Rev. Lett.}\ }\textbf {\bibinfo {volume} {95}},\ \bibinfo
  {pages} {182301} (\bibinfo {year} {2005})},\ \Eprint
  {http://arxiv.org/abs/nucl-th/0505052} {arXiv:nucl-th/0505052 [nucl-th]}
  \BibitemShut {NoStop}%
\bibitem [{\citenamefont {Haussler}\ \emph {et~al.}(2006)\citenamefont
  {Haussler}, \citenamefont {Stoecker},\ and\ \citenamefont
  {Bleicher}}]{Haussler:2005ei}%
  \BibitemOpen
  \bibfield  {author} {\bibinfo {author} {\bibfnamefont {S.}~\bibnamefont
  {Haussler}}, \bibinfo {author} {\bibfnamefont {H.}~\bibnamefont {Stoecker}},
  \ and\ \bibinfo {author} {\bibfnamefont {M.}~\bibnamefont {Bleicher}},\
  }\href {\doibase 10.1103/PhysRevC.73.021901} {\bibfield  {journal} {\bibinfo
  {journal} {Phys. Rev.}\ }\textbf {\bibinfo {volume} {C73}},\ \bibinfo {pages}
  {021901} (\bibinfo {year} {2006})},\ \Eprint
  {http://arxiv.org/abs/hep-ph/0507189} {arXiv:hep-ph/0507189 [hep-ph]}
  \BibitemShut {NoStop}%
\bibitem [{\citenamefont {Cheng}\ \emph {et~al.}(2009)\citenamefont {Cheng}
  \emph {et~al.}}]{Cheng:2008zh}%
  \BibitemOpen
  \bibfield  {author} {\bibinfo {author} {\bibfnamefont {M.}~\bibnamefont
  {Cheng}} \emph {et~al.},\ }\href {\doibase 10.1103/PhysRevD.79.074505}
  {\bibfield  {journal} {\bibinfo  {journal} {Phys. Rev.}\ }\textbf {\bibinfo
  {volume} {D79}},\ \bibinfo {pages} {074505} (\bibinfo {year} {2009})},\
  \Eprint {http://arxiv.org/abs/0811.1006} {arXiv:0811.1006 [hep-lat]}
  \BibitemShut {NoStop}%
\bibitem [{\citenamefont {Zhang}\ \emph {et~al.}(2010)\citenamefont {Zhang},
  \citenamefont {Chen}, \citenamefont {Crawford}, \citenamefont {Keane},
  \citenamefont {Ma},\ and\ \citenamefont {Xu}}]{Zhang:2009ba}%
  \BibitemOpen
  \bibfield  {author} {\bibinfo {author} {\bibfnamefont {S.}~\bibnamefont
  {Zhang}}, \bibinfo {author} {\bibfnamefont {J.~H.}\ \bibnamefont {Chen}},
  \bibinfo {author} {\bibfnamefont {H.}~\bibnamefont {Crawford}}, \bibinfo
  {author} {\bibfnamefont {D.}~\bibnamefont {Keane}}, \bibinfo {author}
  {\bibfnamefont {Y.~G.}\ \bibnamefont {Ma}}, \ and\ \bibinfo {author}
  {\bibfnamefont {Z.~B.}\ \bibnamefont {Xu}},\ }\href {\doibase
  10.1016/j.physletb.2010.01.034} {\bibfield  {journal} {\bibinfo  {journal}
  {Phys. Lett.}\ }\textbf {\bibinfo {volume} {B684}},\ \bibinfo {pages} {224}
  (\bibinfo {year} {2010})},\ \Eprint {http://arxiv.org/abs/0908.3357}
  {arXiv:0908.3357 [nucl-ex]} \BibitemShut {NoStop}%
\bibitem [{\citenamefont {Steinheimer}\ \emph {et~al.}(2012)\citenamefont
  {Steinheimer}, \citenamefont {Gudima}, \citenamefont {Botvina}, \citenamefont
  {Mishustin}, \citenamefont {Bleicher},\ and\ \citenamefont
  {Stocker}}]{Steinheimer:2012tb}%
  \BibitemOpen
  \bibfield  {author} {\bibinfo {author} {\bibfnamefont {J.}~\bibnamefont
  {Steinheimer}}, \bibinfo {author} {\bibfnamefont {K.}~\bibnamefont {Gudima}},
  \bibinfo {author} {\bibfnamefont {A.}~\bibnamefont {Botvina}}, \bibinfo
  {author} {\bibfnamefont {I.}~\bibnamefont {Mishustin}}, \bibinfo {author}
  {\bibfnamefont {M.}~\bibnamefont {Bleicher}}, \ and\ \bibinfo {author}
  {\bibfnamefont {H.}~\bibnamefont {Stocker}},\ }\href {\doibase
  10.1016/j.physletb.2012.06.069} {\bibfield  {journal} {\bibinfo  {journal}
  {Phys. Lett.}\ }\textbf {\bibinfo {volume} {B714}},\ \bibinfo {pages} {85}
  (\bibinfo {year} {2012})},\ \Eprint {http://arxiv.org/abs/1203.2547}
  {arXiv:1203.2547 [nucl-th]} \BibitemShut {NoStop}%
\bibitem [{\citenamefont {Armstrong}\ \emph {et~al.}(2004)\citenamefont
  {Armstrong} \emph {et~al.}}]{Armstrong:2002xh}%
  \BibitemOpen
  \bibfield  {author} {\bibinfo {author} {\bibfnamefont {T.~A.}\ \bibnamefont
  {Armstrong}} \emph {et~al.} (\bibinfo {collaboration} {E864}),\ }\href
  {\doibase 10.1103/PhysRevC.70.024902} {\bibfield  {journal} {\bibinfo
  {journal} {Phys. Rev.}\ }\textbf {\bibinfo {volume} {C70}},\ \bibinfo {pages}
  {024902} (\bibinfo {year} {2004})},\ \Eprint
  {http://arxiv.org/abs/nucl-ex/0211010} {arXiv:nucl-ex/0211010 [nucl-ex]}
  \BibitemShut {NoStop}%
\bibitem [{\citenamefont {Abelev}\ \emph {et~al.}(2010)\citenamefont {Abelev}
  \emph {et~al.}}]{Abelev:2010rv}%
  \BibitemOpen
  \bibfield  {author} {\bibinfo {author} {\bibfnamefont {B.~I.}\ \bibnamefont
  {Abelev}} \emph {et~al.} (\bibinfo {collaboration} {STAR}),\ }\href {\doibase
  10.1126/science.1183980} {\bibfield  {journal} {\bibinfo  {journal}
  {Science}\ }\textbf {\bibinfo {volume} {328}},\ \bibinfo {pages} {58}
  (\bibinfo {year} {2010})},\ \Eprint {http://arxiv.org/abs/1003.2030}
  {arXiv:1003.2030 [nucl-ex]} \BibitemShut {NoStop}%
\bibitem [{\citenamefont {Adam}\ \emph
  {et~al.}(2016{\natexlab{a}})\citenamefont {Adam} \emph
  {et~al.}}]{Adam:2015yta}%
  \BibitemOpen
  \bibfield  {author} {\bibinfo {author} {\bibfnamefont {J.}~\bibnamefont
  {Adam}} \emph {et~al.} (\bibinfo {collaboration} {ALICE}),\ }\href {\doibase
  10.1016/j.physletb.2016.01.040} {\bibfield  {journal} {\bibinfo  {journal}
  {Phys. Lett.}\ }\textbf {\bibinfo {volume} {B754}},\ \bibinfo {pages} {360}
  (\bibinfo {year} {2016}{\natexlab{a}})},\ \Eprint
  {http://arxiv.org/abs/1506.08453} {arXiv:1506.08453 [nucl-ex]} \BibitemShut
  {NoStop}%
\bibitem [{\citenamefont {Andronic}\ \emph {et~al.}(2011)\citenamefont
  {Andronic}, \citenamefont {Braun-Munzinger}, \citenamefont {Stachel},\ and\
  \citenamefont {Stocker}}]{Andronic:2010qu}%
  \BibitemOpen
  \bibfield  {author} {\bibinfo {author} {\bibfnamefont {A.}~\bibnamefont
  {Andronic}}, \bibinfo {author} {\bibfnamefont {P.}~\bibnamefont
  {Braun-Munzinger}}, \bibinfo {author} {\bibfnamefont {J.}~\bibnamefont
  {Stachel}}, \ and\ \bibinfo {author} {\bibfnamefont {H.}~\bibnamefont
  {Stocker}},\ }\href {\doibase 10.1016/j.physletb.2011.01.053} {\bibfield
  {journal} {\bibinfo  {journal} {Phys. Lett.}\ }\textbf {\bibinfo {volume}
  {B697}},\ \bibinfo {pages} {203} (\bibinfo {year} {2011})},\ \Eprint
  {http://arxiv.org/abs/1010.2995} {arXiv:1010.2995 [nucl-th]} \BibitemShut
  {NoStop}%
\bibitem [{\citenamefont {Cleymans}\ \emph {et~al.}(2011)\citenamefont
  {Cleymans}, \citenamefont {Kabana}, \citenamefont {Kraus}, \citenamefont
  {Oeschler}, \citenamefont {Redlich},\ and\ \citenamefont
  {Sharma}}]{Cleymans:2011pe}%
  \BibitemOpen
  \bibfield  {author} {\bibinfo {author} {\bibfnamefont {J.}~\bibnamefont
  {Cleymans}}, \bibinfo {author} {\bibfnamefont {S.}~\bibnamefont {Kabana}},
  \bibinfo {author} {\bibfnamefont {I.}~\bibnamefont {Kraus}}, \bibinfo
  {author} {\bibfnamefont {H.}~\bibnamefont {Oeschler}}, \bibinfo {author}
  {\bibfnamefont {K.}~\bibnamefont {Redlich}}, \ and\ \bibinfo {author}
  {\bibfnamefont {N.}~\bibnamefont {Sharma}},\ }\href {\doibase
  10.1103/PhysRevC.84.054916} {\bibfield  {journal} {\bibinfo  {journal} {Phys.
  Rev.}\ }\textbf {\bibinfo {volume} {C84}},\ \bibinfo {pages} {054916}
  (\bibinfo {year} {2011})},\ \Eprint {http://arxiv.org/abs/1105.3719}
  {arXiv:1105.3719 [hep-ph]} \BibitemShut {NoStop}%
\bibitem [{\citenamefont {Sun}\ and\ \citenamefont {Chen}(2016)}]{Sun:2015ulc}%
  \BibitemOpen
  \bibfield  {author} {\bibinfo {author} {\bibfnamefont {K.-J.}\ \bibnamefont
  {Sun}}\ and\ \bibinfo {author} {\bibfnamefont {L.-W.}\ \bibnamefont {Chen}},\
  }\href {\doibase 10.1103/PhysRevC.93.064909} {\bibfield  {journal} {\bibinfo
  {journal} {Phys. Rev.}\ }\textbf {\bibinfo {volume} {C93}},\ \bibinfo {pages}
  {064909} (\bibinfo {year} {2016})},\ \Eprint
  {http://arxiv.org/abs/1512.00692} {arXiv:1512.00692 [nucl-th]} \BibitemShut
  {NoStop}%
\bibitem [{\citenamefont {Sun}\ \emph {et~al.}(2019)\citenamefont {Sun},
  \citenamefont {Ko},\ and\ \citenamefont {Donigus}}]{Sun:2018mqq}%
  \BibitemOpen
  \bibfield  {author} {\bibinfo {author} {\bibfnamefont {K.-J.}\ \bibnamefont
  {Sun}}, \bibinfo {author} {\bibfnamefont {C.~M.}\ \bibnamefont {Ko}}, \ and\
  \bibinfo {author} {\bibfnamefont {B.}~\bibnamefont {Donigus}},\ }\href
  {\doibase 10.1016/j.physletb.2019.03.033} {\bibfield  {journal} {\bibinfo
  {journal} {Phys. Lett.}\ }\textbf {\bibinfo {volume} {B792}},\ \bibinfo
  {pages} {132} (\bibinfo {year} {2019})},\ \Eprint
  {http://arxiv.org/abs/1812.05175} {arXiv:1812.05175 [nucl-th]} \BibitemShut
  {NoStop}%
\bibitem [{\citenamefont {Kerman}\ and\ \citenamefont
  {Weiss}(1973)}]{Kerman1973}%
  \BibitemOpen
  \bibfield  {author} {\bibinfo {author} {\bibfnamefont {A.~K.}\ \bibnamefont
  {Kerman}}\ and\ \bibinfo {author} {\bibfnamefont {M.~S.}\ \bibnamefont
  {Weiss}},\ }\href@noop {} {\bibfield  {journal} {\bibinfo  {journal} {Phys.
  Rev. C}\ }\textbf {\bibinfo {volume} {8}},\ \bibinfo {pages} {408} (\bibinfo
  {year} {1973})}\BibitemShut {NoStop}%
\bibitem [{\citenamefont {Ko}(1985)}]{PhysRevC.32.326}%
  \BibitemOpen
  \bibfield  {author} {\bibinfo {author} {\bibfnamefont {C.~M.}\ \bibnamefont
  {Ko}},\ }\href {\doibase 10.1103/PhysRevC.32.326} {\bibfield  {journal}
  {\bibinfo  {journal} {Phys. Rev. C}\ }\textbf {\bibinfo {volume} {32}},\
  \bibinfo {pages} {326} (\bibinfo {year} {1985})}\BibitemShut {NoStop}%
\bibitem [{\citenamefont {Rappold}\ \emph {et~al.}(2013)\citenamefont {Rappold}
  \emph {et~al.}}]{2013.NPA.913.170}%
  \BibitemOpen
  \bibfield  {author} {\bibinfo {author} {\bibfnamefont {C.}~\bibnamefont
  {Rappold}} \emph {et~al.},\ }\href@noop {} {\bibfield  {journal} {\bibinfo
  {journal} {Nucl. Phys. A}\ }\textbf {\bibinfo {volume} {913}},\ \bibinfo
  {pages} {170} (\bibinfo {year} {2013})}\BibitemShut {NoStop}%
\bibitem [{\citenamefont {Liu}\ \emph {et~al.}(2017)\citenamefont {Liu},
  \citenamefont {Chen}, \citenamefont {Ma},\ and\ \citenamefont
  {Zhang}}]{Liu:2017rjm}%
  \BibitemOpen
  \bibfield  {author} {\bibinfo {author} {\bibfnamefont {P.}~\bibnamefont
  {Liu}}, \bibinfo {author} {\bibfnamefont {J.}~\bibnamefont {Chen}}, \bibinfo
  {author} {\bibfnamefont {Y.-G.}\ \bibnamefont {Ma}}, \ and\ \bibinfo {author}
  {\bibfnamefont {S.}~\bibnamefont {Zhang}},\ }\href {\doibase
  10.1007/s41365-017-0207-x} {\bibfield  {journal} {\bibinfo  {journal} {Nucl.\
  Sci.\ Tech.}\ }\textbf {\bibinfo {volume} {28}},\ \bibinfo {pages} {55}
  (\bibinfo {year} {2017})},\ \bibinfo {note} {[Erratum: Nucl.Sci.Tech. 28, 89
  (2017)]}\BibitemShut {NoStop}%
\bibitem [{\citenamefont {Zhang}\ and\ \citenamefont
  {Ko}(2018)}]{Zhang:2018euf}%
  \BibitemOpen
  \bibfield  {author} {\bibinfo {author} {\bibfnamefont {Z.}~\bibnamefont
  {Zhang}}\ and\ \bibinfo {author} {\bibfnamefont {C.~M.}\ \bibnamefont {Ko}},\
  }\href {\doibase 10.1016/j.physletb.2018.03.003} {\bibfield  {journal}
  {\bibinfo  {journal} {Phys. Lett.}\ }\textbf {\bibinfo {volume} {B780}},\
  \bibinfo {pages} {191} (\bibinfo {year} {2018})}\BibitemShut {NoStop}%
\bibitem [{\citenamefont {Gal}\ \emph {et~al.}(2016)\citenamefont {Gal},
  \citenamefont {Hungerford},\ and\ \citenamefont
  {Millener}}]{2016.RMP.88.035004}%
  \BibitemOpen
  \bibfield  {author} {\bibinfo {author} {\bibfnamefont {A.}~\bibnamefont
  {Gal}}, \bibinfo {author} {\bibfnamefont {E.~V.}\ \bibnamefont {Hungerford}},
  \ and\ \bibinfo {author} {\bibfnamefont {D.~J.}\ \bibnamefont {Millener}},\
  }\href@noop {} {\bibfield  {journal} {\bibinfo  {journal} {Rev. Mod. Phys.}\
  }\textbf {\bibinfo {volume} {88}},\ \bibinfo {pages} {035004} (\bibinfo
  {year} {2016})}\BibitemShut {NoStop}%
\bibitem [{\citenamefont {Chen}\ \emph {et~al.}(2018)\citenamefont {Chen},
  \citenamefont {Keane}, \citenamefont {Ma}, \citenamefont {Tang},\ and\
  \citenamefont {Xu}}]{CHEN20181}%
  \BibitemOpen
  \bibfield  {author} {\bibinfo {author} {\bibfnamefont {J.}~\bibnamefont
  {Chen}}, \bibinfo {author} {\bibfnamefont {D.}~\bibnamefont {Keane}},
  \bibinfo {author} {\bibfnamefont {Y.-G.}\ \bibnamefont {Ma}}, \bibinfo
  {author} {\bibfnamefont {A.}~\bibnamefont {Tang}}, \ and\ \bibinfo {author}
  {\bibfnamefont {Z.}~\bibnamefont {Xu}},\ }\href {\doibase
  https://doi.org/10.1016/j.physrep.2018.07.002} {\bibfield  {journal}
  {\bibinfo  {journal} {Phys. Rept.}\ }\textbf {\bibinfo {volume} {760}},\
  \bibinfo {pages} {1 } (\bibinfo {year} {2018})},\ \Eprint
  {http://arxiv.org/abs/1808.09619} {arXiv:1808.09619 [nucl-ex]} \BibitemShut
  {NoStop}%
\bibitem [{\citenamefont {Braun-Munzinger}\ and\ \citenamefont
  {Donigus}(2019)}]{Braun-Munzinger:2018hat}%
  \BibitemOpen
  \bibfield  {author} {\bibinfo {author} {\bibfnamefont {P.}~\bibnamefont
  {Braun-Munzinger}}\ and\ \bibinfo {author} {\bibfnamefont {B.}~\bibnamefont
  {Donigus}},\ }\href {\doibase 10.1016/j.nuclphysa.2019.02.006} {\bibfield
  {journal} {\bibinfo  {journal} {Nucl. Phys.}\ }\textbf {\bibinfo {volume}
  {A987}},\ \bibinfo {pages} {144} (\bibinfo {year} {2019})},\ \Eprint
  {http://arxiv.org/abs/1809.04681} {arXiv:1809.04681 [nucl-ex]} \BibitemShut
  {NoStop}%
\bibitem [{\citenamefont {Davis}(2005)}]{Davis:2005}%
  \BibitemOpen
  \bibfield  {author} {\bibinfo {author} {\bibfnamefont {D.}~\bibnamefont
  {Davis}},\ }\href {\doibase doi:10.1016/j.nuclphysa.2005.01.002} {\bibfield
  {journal} {\bibinfo  {journal} {Nucl. Phys.}\ }\textbf {\bibinfo {volume}
  {A754}},\ \bibinfo {pages} {3c} (\bibinfo {year} {2005})},\ \bibinfo {note}
  {the HYP 2003}\BibitemShut {NoStop}%
\bibitem [{\citenamefont {Adam}\ \emph
  {et~al.}(2019{\natexlab{a}})\citenamefont {Adam} \emph
  {et~al.}}]{Adam:2019phl}%
  \BibitemOpen
  \bibfield  {author} {\bibinfo {author} {\bibfnamefont {J.}~\bibnamefont
  {Adam}} \emph {et~al.} (\bibinfo {collaboration} {STAR}),\ }\href@noop {} {\
  (\bibinfo {year} {2019}{\natexlab{a}})},\ \Eprint
  {http://arxiv.org/abs/1904.10520} {arXiv:1904.10520 [hep-ex]} \BibitemShut
  {NoStop}%
\bibitem [{\citenamefont {Liu}\ \emph {et~al.}(2019)\citenamefont {Liu},
  \citenamefont {Chen}, \citenamefont {Keane}, \citenamefont {Xu},\ and\
  \citenamefont {Ma}}]{Liu:2019mlm}%
  \BibitemOpen
  \bibfield  {author} {\bibinfo {author} {\bibfnamefont {P.}~\bibnamefont
  {Liu}}, \bibinfo {author} {\bibfnamefont {J.}~\bibnamefont {Chen}}, \bibinfo
  {author} {\bibfnamefont {D.}~\bibnamefont {Keane}}, \bibinfo {author}
  {\bibfnamefont {Z.}~\bibnamefont {Xu}}, \ and\ \bibinfo {author}
  {\bibfnamefont {Y.-G.}\ \bibnamefont {Ma}},\ }\href {\doibase
  10.1088/1674-1137/43/12/124001} {\bibfield  {journal} {\bibinfo  {journal}
  {Chin. Phys.}\ }\textbf {\bibinfo {volume} {C43}},\ \bibinfo {pages} {124001}
  (\bibinfo {year} {2019})},\ \Eprint {http://arxiv.org/abs/1908.03134}
  {arXiv:1908.03134 [nucl-ex]} \BibitemShut {NoStop}%
\bibitem [{\citenamefont {Zhu}(2013)}]{ZHU2013551c}%
  \BibitemOpen
  \bibfield  {author} {\bibinfo {author} {\bibfnamefont {Y.}~\bibnamefont
  {Zhu}} (\bibinfo {collaboration} {STAR}),\ }\href {\doibase
  https://doi.org/10.1016/j.nuclphysa.2013.02.074} {\bibfield  {journal}
  {\bibinfo  {journal} {Nucl. Phys.}\ }\textbf {\bibinfo {volume} {A904-905}},\
  \bibinfo {pages} {551c } (\bibinfo {year} {2013})},\ \bibinfo {note} {the
  Quark Matter 2012}\BibitemShut {NoStop}%
\bibitem [{\citenamefont {Adler}\ \emph {et~al.}(2004)\citenamefont {Adler}
  \emph {et~al.}}]{Adler:2003cb}%
  \BibitemOpen
  \bibfield  {author} {\bibinfo {author} {\bibfnamefont {S.~S.}\ \bibnamefont
  {Adler}} \emph {et~al.} (\bibinfo {collaboration} {PHENIX}),\ }\href
  {\doibase 10.1103/PhysRevC.69.034909} {\bibfield  {journal} {\bibinfo
  {journal} {Phys. Rev.}\ }\textbf {\bibinfo {volume} {C69}},\ \bibinfo {pages}
  {034909} (\bibinfo {year} {2004})},\ \Eprint
  {http://arxiv.org/abs/nucl-ex/0307022} {arXiv:nucl-ex/0307022 [nucl-ex]}
  \BibitemShut {NoStop}%
\bibitem [{\citenamefont {Adamczyk}\ \emph {et~al.}(2018)\citenamefont
  {Adamczyk} \emph {et~al.}}]{Adamczyk:2017buv}%
  \BibitemOpen
  \bibfield  {author} {\bibinfo {author} {\bibfnamefont {L.}~\bibnamefont
  {Adamczyk}} \emph {et~al.} (\bibinfo {collaboration} {STAR}),\ }\href
  {\doibase 10.1103/PhysRevC.97.054909} {\bibfield  {journal} {\bibinfo
  {journal} {Phys. Rev.}\ }\textbf {\bibinfo {volume} {C97}},\ \bibinfo {pages}
  {054909} (\bibinfo {year} {2018})},\ \Eprint
  {http://arxiv.org/abs/1710.00436} {arXiv:1710.00436 [nucl-ex]} \BibitemShut
  {NoStop}%
\bibitem [{\citenamefont {Sun}\ \emph {et~al.}(2018)\citenamefont {Sun},
  \citenamefont {Chen}, \citenamefont {Ko}, \citenamefont {Pu},\ and\
  \citenamefont {Xu}}]{Sun:2018jhg}%
  \BibitemOpen
  \bibfield  {author} {\bibinfo {author} {\bibfnamefont {K.-J.}\ \bibnamefont
  {Sun}}, \bibinfo {author} {\bibfnamefont {L.-W.}\ \bibnamefont {Chen}},
  \bibinfo {author} {\bibfnamefont {C.~M.}\ \bibnamefont {Ko}}, \bibinfo
  {author} {\bibfnamefont {J.}~\bibnamefont {Pu}}, \ and\ \bibinfo {author}
  {\bibfnamefont {Z.}~\bibnamefont {Xu}},\ }\href {\doibase
  10.1016/j.physletb.2018.04.035} {\bibfield  {journal} {\bibinfo  {journal}
  {Phys. Lett.}\ }\textbf {\bibinfo {volume} {B781}},\ \bibinfo {pages} {499}
  (\bibinfo {year} {2018})},\ \Eprint {http://arxiv.org/abs/1801.09382}
  {arXiv:1801.09382 [nucl-th]} \BibitemShut {NoStop}%
\bibitem [{\citenamefont {Ko}\ and\ \citenamefont {Li}(2016)}]{Ko:2016ioz}%
  \BibitemOpen
  \bibfield  {author} {\bibinfo {author} {\bibfnamefont {C.~M.}\ \bibnamefont
  {Ko}}\ and\ \bibinfo {author} {\bibfnamefont {F.}~\bibnamefont {Li}},\ }\href
  {\doibase 10.1007/s41365-016-0141-3} {\bibfield  {journal} {\bibinfo
  {journal} {Nucl. Sci. Tech.}\ }\textbf {\bibinfo {volume} {27}},\ \bibinfo
  {pages} {140} (\bibinfo {year} {2016})},\ \bibinfo {note} {the IWND
  2016}\BibitemShut {NoStop}%
\bibitem [{\citenamefont {Shao}\ \emph {et~al.}(2020)\citenamefont {Shao},
  \citenamefont {Chen}, \citenamefont {Ko},\ and\ \citenamefont
  {Sun}}]{Shao:2019xpj}%
  \BibitemOpen
  \bibfield  {author} {\bibinfo {author} {\bibfnamefont {T.}~\bibnamefont
  {Shao}}, \bibinfo {author} {\bibfnamefont {J.}~\bibnamefont {Chen}}, \bibinfo
  {author} {\bibfnamefont {C.~M.}\ \bibnamefont {Ko}}, \ and\ \bibinfo {author}
  {\bibfnamefont {K.-J.}\ \bibnamefont {Sun}},\ }\href {\doibase
  10.1016/j.physletb.2019.135177} {\bibfield  {journal} {\bibinfo  {journal}
  {Phys. Lett.}\ }\textbf {\bibinfo {volume} {B801}},\ \bibinfo {pages}
  {135177} (\bibinfo {year} {2020})},\ \Eprint
  {http://arxiv.org/abs/1910.14281} {arXiv:1910.14281 [hep-ph]} \BibitemShut
  {NoStop}%
\bibitem [{\citenamefont {Jin}\ \emph {et~al.}(2019)\citenamefont {Jin},
  \citenamefont {Chen}, \citenamefont {Lin}, \citenamefont {Ma}, \citenamefont
  {Ma},\ and\ \citenamefont {Zhang}}]{Jin:2019}%
  \BibitemOpen
  \bibfield  {author} {\bibinfo {author} {\bibfnamefont {X.}~\bibnamefont
  {Jin}}, \bibinfo {author} {\bibfnamefont {J.}~\bibnamefont {Chen}}, \bibinfo
  {author} {\bibfnamefont {Z.}~\bibnamefont {Lin}}, \bibinfo {author}
  {\bibfnamefont {G.}~\bibnamefont {Ma}}, \bibinfo {author} {\bibfnamefont
  {Y.}~\bibnamefont {Ma}}, \ and\ \bibinfo {author} {\bibfnamefont
  {S.}~\bibnamefont {Zhang}},\ }\href {\doibase 10.1007/s11433-018-9272-4}
  {\bibfield  {journal} {\bibinfo  {journal} {Sci. China-Phys. Mech. Astron.}\
  }\textbf {\bibinfo {volume} {62}},\ \bibinfo {pages} {011012} (\bibinfo
  {year} {2019})}\BibitemShut {NoStop}%
\bibitem [{\citenamefont {Sun}\ \emph {et~al.}(2017)\citenamefont {Sun},
  \citenamefont {Chen}, \citenamefont {Ko},\ and\ \citenamefont
  {Xu}}]{Sun:2017xrx}%
  \BibitemOpen
  \bibfield  {author} {\bibinfo {author} {\bibfnamefont {K.-J.}\ \bibnamefont
  {Sun}}, \bibinfo {author} {\bibfnamefont {L.-W.}\ \bibnamefont {Chen}},
  \bibinfo {author} {\bibfnamefont {C.~M.}\ \bibnamefont {Ko}}, \ and\ \bibinfo
  {author} {\bibfnamefont {Z.}~\bibnamefont {Xu}},\ }\href {\doibase
  10.1016/j.physletb.2017.09.056} {\bibfield  {journal} {\bibinfo  {journal}
  {Phys.\ Lett.\ B}\ }\textbf {\bibinfo {volume} {774}},\ \bibinfo {pages}
  {103} (\bibinfo {year} {2017})},\ \Eprint {http://arxiv.org/abs/1702.07620}
  {arXiv:1702.07620 [nucl-th]} \BibitemShut {NoStop}%
\bibitem [{\citenamefont {Adamczyk}\ \emph {et~al.}(2017)\citenamefont
  {Adamczyk} \emph {et~al.}}]{Adamczyk:2017iwn}%
  \BibitemOpen
  \bibfield  {author} {\bibinfo {author} {\bibfnamefont {L.}~\bibnamefont
  {Adamczyk}} \emph {et~al.} (\bibinfo {collaboration} {STAR}),\ }\href
  {\doibase 10.1103/PhysRevC.96.044904} {\bibfield  {journal} {\bibinfo
  {journal} {Phys. Rev.}\ }\textbf {\bibinfo {volume} {C96}},\ \bibinfo {pages}
  {044904} (\bibinfo {year} {2017})},\ \Eprint
  {http://arxiv.org/abs/1701.07065} {arXiv:1701.07065 [nucl-ex]} \BibitemShut
  {NoStop}%
\bibitem [{\citenamefont {Abelev}\ \emph {et~al.}(2009)\citenamefont {Abelev}
  \emph {et~al.}}]{Abelev:2008ab}%
  \BibitemOpen
  \bibfield  {author} {\bibinfo {author} {\bibfnamefont {B.~I.}\ \bibnamefont
  {Abelev}} \emph {et~al.} (\bibinfo {collaboration} {STAR}),\ }\href {\doibase
  10.1103/PhysRevC.79.034909} {\bibfield  {journal} {\bibinfo  {journal} {Phys.
  Rev.}\ }\textbf {\bibinfo {volume} {C79}},\ \bibinfo {pages} {034909}
  (\bibinfo {year} {2009})},\ \Eprint {http://arxiv.org/abs/0808.2041}
  {arXiv:0808.2041 [nucl-ex]} \BibitemShut {NoStop}%
\bibitem [{\citenamefont {Andronic}\ \emph {et~al.}(2006)\citenamefont
  {Andronic}, \citenamefont {Braun-Munzinger},\ and\ \citenamefont
  {Stachel}}]{Andronic:2005yp}%
  \BibitemOpen
  \bibfield  {author} {\bibinfo {author} {\bibfnamefont {A.}~\bibnamefont
  {Andronic}}, \bibinfo {author} {\bibfnamefont {P.}~\bibnamefont
  {Braun-Munzinger}}, \ and\ \bibinfo {author} {\bibfnamefont {J.}~\bibnamefont
  {Stachel}},\ }\href {\doibase 10.1016/j.nuclphysa.2006.03.012} {\bibfield
  {journal} {\bibinfo  {journal} {Nucl. Phys.}\ }\textbf {\bibinfo {volume}
  {A772}},\ \bibinfo {pages} {167} (\bibinfo {year} {2006})},\ \Eprint
  {http://arxiv.org/abs/nucl-th/0511071} {arXiv:nucl-th/0511071 [nucl-th]}
  \BibitemShut {NoStop}%
\bibitem [{\citenamefont {Sun}\ and\ \citenamefont {Chen}(2017)}]{Sun:2017ooe}%
  \BibitemOpen
  \bibfield  {author} {\bibinfo {author} {\bibfnamefont {K.-J.}\ \bibnamefont
  {Sun}}\ and\ \bibinfo {author} {\bibfnamefont {L.-W.}\ \bibnamefont {Chen}},\
  }\href {\doibase 10.1103/PhysRevC.95.044905} {\bibfield  {journal} {\bibinfo
  {journal} {Phys. Rev.}\ }\textbf {\bibinfo {volume} {C95}},\ \bibinfo {pages}
  {044905} (\bibinfo {year} {2017})},\ \Eprint
  {http://arxiv.org/abs/1701.01935} {arXiv:1701.01935 [nucl-th]} \BibitemShut
  {NoStop}%
\bibitem [{\citenamefont {Adam}\ \emph
  {et~al.}(2016{\natexlab{b}})\citenamefont {Adam} \emph
  {et~al.}}]{Adam:2015vda}%
  \BibitemOpen
  \bibfield  {author} {\bibinfo {author} {\bibfnamefont {J.}~\bibnamefont
  {Adam}} \emph {et~al.} (\bibinfo {collaboration} {ALICE}),\ }\href {\doibase
  10.1103/PhysRevC.93.024917} {\bibfield  {journal} {\bibinfo  {journal} {Phys.
  Rev.}\ }\textbf {\bibinfo {volume} {C93}},\ \bibinfo {pages} {024917}
  (\bibinfo {year} {2016}{\natexlab{b}})},\ \Eprint
  {http://arxiv.org/abs/1506.08951} {arXiv:1506.08951 [nucl-ex]} \BibitemShut
  {NoStop}%
\bibitem [{\citenamefont {Zhang}(2020)}]{Zhang:2020ewj}%
  \BibitemOpen
  \bibfield  {author} {\bibinfo {author} {\bibfnamefont {D.}~\bibnamefont
  {Zhang}} (\bibinfo {collaboration} {STAR})\ }(\bibinfo {year} {2020})\
  \bibinfo {note} {the Quark Matter 2019},\ \Eprint
  {http://arxiv.org/abs/2002.10677} {arXiv:2002.10677 [nucl-ex]} \BibitemShut
  {NoStop}%
\bibitem [{\citenamefont {Armstrong}\ \emph {et~al.}(2000)\citenamefont
  {Armstrong} \emph {et~al.}}]{Armstrong:2000gz}%
  \BibitemOpen
  \bibfield  {author} {\bibinfo {author} {\bibfnamefont {T.~A.}\ \bibnamefont
  {Armstrong}} \emph {et~al.} (\bibinfo {collaboration} {E864}),\ }\href
  {\doibase 10.1103/PhysRevC.61.064908} {\bibfield  {journal} {\bibinfo
  {journal} {Phys. Rev.}\ }\textbf {\bibinfo {volume} {C61}},\ \bibinfo {pages}
  {064908} (\bibinfo {year} {2000})},\ \Eprint
  {http://arxiv.org/abs/nucl-ex/0003009} {arXiv:nucl-ex/0003009 [nucl-ex]}
  \BibitemShut {NoStop}%
\bibitem [{\citenamefont {Albergo}\ \emph {et~al.}(2002)\citenamefont {Albergo}
  \emph {et~al.}}]{Albergo:2002tn}%
  \BibitemOpen
  \bibfield  {author} {\bibinfo {author} {\bibfnamefont {S.}~\bibnamefont
  {Albergo}} \emph {et~al.} (\bibinfo {collaboration} {E896}),\ }\href
  {\doibase 10.1103/PhysRevLett.88.062301} {\bibfield  {journal} {\bibinfo
  {journal} {Phys. Rev. Lett.}\ }\textbf {\bibinfo {volume} {88}},\ \bibinfo
  {pages} {062301} (\bibinfo {year} {2002})}\BibitemShut {NoStop}%
\bibitem [{\citenamefont {Adam}\ \emph
  {et~al.}(2019{\natexlab{b}})\citenamefont {Adam} \emph
  {et~al.}}]{Adam:2019wnb}%
  \BibitemOpen
  \bibfield  {author} {\bibinfo {author} {\bibfnamefont {J.}~\bibnamefont
  {Adam}} \emph {et~al.} (\bibinfo {collaboration} {STAR}),\ }\href {\doibase
  10.1103/PhysRevC.99.064905} {\bibfield  {journal} {\bibinfo  {journal} {Phys.
  Rev.}\ }\textbf {\bibinfo {volume} {C99}},\ \bibinfo {pages} {064905}
  (\bibinfo {year} {2019}{\natexlab{b}})},\ \Eprint
  {http://arxiv.org/abs/1903.11778} {arXiv:1903.11778 [nucl-ex]} \BibitemShut
  {NoStop}%
\bibitem [{\citenamefont {Agakishiev}\ \emph {et~al.}(2012)\citenamefont
  {Agakishiev} \emph {et~al.}}]{Agakishiev:2011ar}%
  \BibitemOpen
  \bibfield  {author} {\bibinfo {author} {\bibfnamefont {G.}~\bibnamefont
  {Agakishiev}} \emph {et~al.} (\bibinfo {collaboration} {STAR}),\ }\href
  {\doibase 10.1103/PhysRevLett.108.072301} {\bibfield  {journal} {\bibinfo
  {journal} {Phys. Rev. Lett.}\ }\textbf {\bibinfo {volume} {108}},\ \bibinfo
  {pages} {072301} (\bibinfo {year} {2012})},\ \Eprint
  {http://arxiv.org/abs/1107.2955} {arXiv:1107.2955 [nucl-ex]} \BibitemShut
  {NoStop}%
\bibitem [{\citenamefont {Abelev}\ \emph {et~al.}(2013)\citenamefont {Abelev}
  \emph {et~al.}}]{Abelev:2013xaa}%
  \BibitemOpen
  \bibfield  {author} {\bibinfo {author} {\bibfnamefont {B.~B.}\ \bibnamefont
  {Abelev}} \emph {et~al.} (\bibinfo {collaboration} {ALICE}),\ }\href
  {\doibase 10.1103/PhysRevLett.111.222301} {\bibfield  {journal} {\bibinfo
  {journal} {Phys. Rev. Lett.}\ }\textbf {\bibinfo {volume} {111}},\ \bibinfo
  {pages} {222301} (\bibinfo {year} {2013})},\ \Eprint
  {http://arxiv.org/abs/1307.5530} {arXiv:1307.5530 [nucl-ex]} \BibitemShut
  {NoStop}%
\bibitem [{\citenamefont {Adam}\ \emph
  {et~al.}(2019{\natexlab{c}})\citenamefont {Adam} \emph
  {et~al.}}]{Adam:2019koz}%
  \BibitemOpen
  \bibfield  {author} {\bibinfo {author} {\bibfnamefont {J.}~\bibnamefont
  {Adam}} \emph {et~al.} (\bibinfo {collaboration} {STAR}),\ }\href@noop {} {\
  (\bibinfo {year} {2019}{\natexlab{c}})},\ \Eprint
  {http://arxiv.org/abs/1906.03732} {arXiv:1906.03732 [nucl-ex]} \BibitemShut
  {NoStop}%
\bibitem [{\citenamefont {Oh}\ and\ \citenamefont {Ko}(2007)}]{Oh:2007vf}%
  \BibitemOpen
  \bibfield  {author} {\bibinfo {author} {\bibfnamefont {Y.}~\bibnamefont
  {Oh}}\ and\ \bibinfo {author} {\bibfnamefont {C.~M.}\ \bibnamefont {Ko}},\
  }\href {\doibase 10.1103/PhysRevC.76.054910} {\bibfield  {journal} {\bibinfo
  {journal} {Phys. Rev.}\ }\textbf {\bibinfo {volume} {C76}},\ \bibinfo {pages}
  {054910} (\bibinfo {year} {2007})},\ \Eprint {http://arxiv.org/abs/0707.3332}
  {arXiv:0707.3332 [nucl-th]} \BibitemShut {NoStop}%
\bibitem [{\citenamefont {Oh}\ \emph {et~al.}(2009)\citenamefont {Oh},
  \citenamefont {Lin},\ and\ \citenamefont {Ko}}]{Oh:2009gx}%
  \BibitemOpen
  \bibfield  {author} {\bibinfo {author} {\bibfnamefont {Y.}~\bibnamefont
  {Oh}}, \bibinfo {author} {\bibfnamefont {Z.-W.}\ \bibnamefont {Lin}}, \ and\
  \bibinfo {author} {\bibfnamefont {C.~M.}\ \bibnamefont {Ko}},\ }\href
  {\doibase 10.1103/PhysRevC.80.064902} {\bibfield  {journal} {\bibinfo
  {journal} {Phys. Rev.}\ }\textbf {\bibinfo {volume} {C80}},\ \bibinfo {pages}
  {064902} (\bibinfo {year} {2009})},\ \Eprint {http://arxiv.org/abs/0910.1977}
  {arXiv:0910.1977 [nucl-th]} \BibitemShut {NoStop}%
\bibitem [{\citenamefont {Oliinychenko}\ \emph {et~al.}(2019)\citenamefont
  {Oliinychenko}, \citenamefont {Pang}, \citenamefont {Elfner},\ and\
  \citenamefont {Koch}}]{Oliinychenko:2018ugs}%
  \BibitemOpen
  \bibfield  {author} {\bibinfo {author} {\bibfnamefont {D.}~\bibnamefont
  {Oliinychenko}}, \bibinfo {author} {\bibfnamefont {L.-G.}\ \bibnamefont
  {Pang}}, \bibinfo {author} {\bibfnamefont {H.}~\bibnamefont {Elfner}}, \ and\
  \bibinfo {author} {\bibfnamefont {V.}~\bibnamefont {Koch}},\ }\href {\doibase
  10.1103/PhysRevC.99.044907} {\bibfield  {journal} {\bibinfo  {journal}
  {Phys.\ Rev.\ C}\ }\textbf {\bibinfo {volume} {99}},\ \bibinfo {pages}
  {044907} (\bibinfo {year} {2019})},\ \Eprint
  {http://arxiv.org/abs/1809.03071} {arXiv:1809.03071 [hep-ph]} \BibitemShut
  {NoStop}%
\bibitem [{\citenamefont {Csernai}\ and\ \citenamefont
  {Kapusta}(1986)}]{Csernai:1986qf}%
  \BibitemOpen
  \bibfield  {author} {\bibinfo {author} {\bibfnamefont {L.~P.}\ \bibnamefont
  {Csernai}}\ and\ \bibinfo {author} {\bibfnamefont {J.~I.}\ \bibnamefont
  {Kapusta}},\ }\href {\doibase 10.1016/0370-1573(86)90031-1} {\bibfield
  {journal} {\bibinfo  {journal} {Phys. Rept.}\ }\textbf {\bibinfo {volume}
  {131}},\ \bibinfo {pages} {223} (\bibinfo {year} {1986})}\BibitemShut
  {NoStop}%
\bibitem [{\citenamefont {Dong}\ \emph {et~al.}(2018)\citenamefont {Dong},
  \citenamefont {Wang}, \citenamefont {Chen}, \citenamefont {She},
  \citenamefont {Yan}, \citenamefont {Liu}, \citenamefont {Zhou},\ and\
  \citenamefont {Sa}}]{Dong2018}%
  \BibitemOpen
  \bibfield  {author} {\bibinfo {author} {\bibfnamefont {Z.-J.}\ \bibnamefont
  {Dong}}, \bibinfo {author} {\bibfnamefont {Q.-Y.}\ \bibnamefont {Wang}},
  \bibinfo {author} {\bibfnamefont {G.}~\bibnamefont {Chen}}, \bibinfo {author}
  {\bibfnamefont {Z.-L.}\ \bibnamefont {She}}, \bibinfo {author} {\bibfnamefont
  {Y.-L.}\ \bibnamefont {Yan}}, \bibinfo {author} {\bibfnamefont {F.-X.}\
  \bibnamefont {Liu}}, \bibinfo {author} {\bibfnamefont {D.-M.}\ \bibnamefont
  {Zhou}}, \ and\ \bibinfo {author} {\bibfnamefont {B.-H.}\ \bibnamefont
  {Sa}},\ }\href {\doibase 10.1140/epja/i2018-12580-8} {\bibfield  {journal}
  {\bibinfo  {journal} {Eur. Phys. J. A}\ }\textbf {\bibinfo {volume} {54}},\
  \bibinfo {pages} {144} (\bibinfo {year} {2018})}\BibitemShut {NoStop}%
\bibitem [{\citenamefont {Zhao}\ \emph {et~al.}(2018)\citenamefont {Zhao},
  \citenamefont {Zhu}, \citenamefont {Zheng}, \citenamefont {Ko},\ and\
  \citenamefont {Song}}]{Zhao:2018lyf}%
  \BibitemOpen
  \bibfield  {author} {\bibinfo {author} {\bibfnamefont {W.}~\bibnamefont
  {Zhao}}, \bibinfo {author} {\bibfnamefont {L.}~\bibnamefont {Zhu}}, \bibinfo
  {author} {\bibfnamefont {H.}~\bibnamefont {Zheng}}, \bibinfo {author}
  {\bibfnamefont {C.~M.}\ \bibnamefont {Ko}}, \ and\ \bibinfo {author}
  {\bibfnamefont {H.}~\bibnamefont {Song}},\ }\href {\doibase
  10.1103/PhysRevC.98.054905} {\bibfield  {journal} {\bibinfo  {journal} {Phys.
  Rev.}\ }\textbf {\bibinfo {volume} {C98}},\ \bibinfo {pages} {054905}
  (\bibinfo {year} {2018})},\ \Eprint {http://arxiv.org/abs/1807.02813}
  {arXiv:1807.02813 [nucl-th]} \BibitemShut {NoStop}%
\end{thebibliography}%

\end{document}